\documentclass{article}
\usepackage{epsfig}
\begin{document}
\title{\bf Geometry of quantum group invariant systems}
\author{Marcelo R. Ubriaco\thanks{Electronic address:ubriaco@ltp.uprrp.edu}}
\date{Laboratory of Theoretical Physics\\University of Puerto Rico\\R\'{\i}o Piedras Campus\\
San Juan\\PR 00931, USA}

\maketitle
\begin{abstract}
Starting with the partition functions for quantum group invariant systems we calculate the 
metric in the two-dimensional space defined  by  the parameters $\beta$ and $\gamma=-\beta\mu$ and the corresponding scalar
curvature for these systems in two and three spatial dimensions. Our results exhibit the details of the anyonic behavior 
of quantum group boson and fermion systems as a function of the fugacity $z$ and the quantum group parameter $q$.
For the case of the quantum group $SU_q(2)$, we compare the stability of these systems with the stability of $SU(2)$ invariant boson and fermion systems. 
\end{abstract}

\section{Introduction}

In the last twenty years there has been considerable interest on applications of quantum groups
\cite{Jimbo}\cite{Ch}, in addition to the theory of integrable models, to diverse areas of theoretical physics.
The vast published literature on this subject includes formulations of quantum group versions of Lorentz and Poincar'{e}
algebras \cite{CW}, its use as internal quantum symmetries in quantum mechanics and field theories \cite{AV}, molecular
and nuclear physics \cite{I}, and  the formulation
and study on the implications of imposing  quantum group invariance in thermodynamic systems \cite{MRU1}. In particular, it has been shown \cite{MRU5} that quantum group gases exhibit anyonic behavior in two and thrre dimensions. More recent
applications include gravity theories wherein the discreteness and non-commutative properties 
of space-time  at the Planck scale are approached mathematically  by replacing the local symmetry by a quantum group
symmetry \cite{MN} and applications to the phenomena of  entanglement \cite{KWL}.  

A very different research field  
is the one based on the idea of using geometry to study some properties of thermodynamic systems \cite{Ti}-\cite{AN}.
In particular, it consists in defining a metric in a two dimensional parameter space y calculate the corresponding scalar curvature
as a measure of the correlations strength of the system \cite{IJKK}-\cite{R}, with applications to classical and quantum gases
\cite{R1}\cite{NS}\cite{JM1}\cite{BH}, magnetic systems \cite{JM2}-\cite{JJK}, non-extensive statistical mechanics \cite{T-B}-\cite{O},
anyon gas \cite{MH1}-\cite{MH2}, fractional statistics \cite{MH3}, deformed boson and fermion systems \cite{MH4} and systems
with fractal distribution functions \cite{MRU}.  Some of the basic 
 results of these approaches include the relationships between the metric  
 with the correlations of the  stochastic variables, and  the scalar curvature $R$ 
 with the stability of the system, and the facts that  the scalar curvature $R$ vanishes for the
 classical ideal gas,   $R>0 (R<0)$ for a boson (fermion) ideal gas, and it is singular at a critical point.
Here, we wish to study the geometry of boson and fermion systems with quantum group symmetry. For reasons of simplicity we will consider
the simplest quantum group, $SU_q(2)$, invariant Hamiltonian, and calculate the metric and the corresponding scalar curvature $R$.
This approach will allow us  to obtain information about the stability of quantum group boson (QGB) and quantum group fermion (QGF) systems
as a consequence of the interactions imposed by the requirement of quantum group invariance, and it will also give
us a complete picture about the anyonic behavior of these systems in the whole range of values of the fugacity $z$.
In Section \ref{QGBF} we describe briefly
the formalism that leads to a $SU_q(2)$ invariant boson and fermion Hamiltonians already studied in Ref. \cite{MRU1}. In Section \ref{Scalar}
we calculate the scalar curvature $R$ for the cases of quantum group bosons and fermions in two and three dimensions. In Section \ref{Conc} we discuss and summarize our results.

\section{Quantum Group Boson and Fermions}\label{QGBF}
The quantum group $SU_q(N)$ invariant algebra is given by the 
 following relations            
\begin{equation}
\Omega_j\overline{\Omega}_i=\delta_{ij}\pm q^{\pm1}{\cal R}_{kijl}
\overline{\Omega}_l\Omega_k \label{c1}
\end{equation}
\begin{equation}
\Omega_l\Omega_k=\pm q^{\mp 1}{\cal R}_{jikl}\Omega_j\Omega_i,\label{c2}
\end{equation}
where $\Omega=\Phi,\Psi$ and the upper (lower) sign applies to QGB $\Phi_i$
(QGF $\Psi_i$) operators.  The $N^2\times N^2$ matrix 
 ${\cal R}_{jikl}$ is explicitly written \cite{WZ}
\begin{equation}
{\cal R}_{jikl}=\delta_{jk}\delta_{il}(1+(q-1)\delta_{ij})
+(q-q^{-1})\delta_{ik}\delta_{jl}\theta(j-i),
\end{equation}
where $\theta(j-i)=1$ for $j>i$ and zero otherwise. The new fields  $\Omega_i'=\sum_{i=1}
^{N}T_{ij}\Omega_j$, the
$SU_{q}(N)$ transformation matrix $T$ 
and the $R$-matrix  
satisfy the well known algebraic relations \cite{Ta}
\begin{equation}
{\cal R}T_1T_2=T_2T_1{\cal R},\label{T}
\end{equation}
and
\begin{equation}
{\cal R}_{12}{\cal R}_{13}{\cal R}_{23}={\cal R}_{23}{\cal R}_{13}{\cal R}_{12},
\end{equation}
with the standard embedding $T_1=T\otimes 1$, $T_2=1\otimes T$
$\in V\otimes V$ and $({\cal R}_{23})_{ijk,i'j'k'}=
\delta_{ii'} {\cal R}_{jk,j'k'} \in V\otimes V\otimes V$.

In particular, the quantum group  $SU_q(2)$ consists of the set of matrices
$T=\left(\begin{array}{cc} a & b \\ c & d\end{array}\right)$
with elements $\{a,b,c,d\}$ generating the algebra
\begin{eqnarray}
ab=q^{-1}ba  & , & ac=q^{-1}ca \nonumber \\
bc=cb & , & dc=qcd  \nonumber \\
db=qbd & , &  da-ad=(q-q^{-1})bc  \nonumber \\
& & det_{q}T\equiv ad-q^{-1}bc=1 ,
\end{eqnarray} 
with the unitary conditions \cite{VWZ} $\overline{a}=d, \overline{b}=q^{-1}c$
and $q\in {\bf R}$. Hereafter, we take $0\leq q<\infty$.
The  corresponding $\cal R$ matrix is

$$\cal R=\left(\begin{array}{cccc} q & 0 & 0 & 0 \\ 0 & 1 & 0 & 0 \\
0 & q-q^{-1} & 1 & 0 \\ 0 & 0 & 0 & q\end{array}\right).$$
For $q=1$. Equations (\ref {c1}) and (\ref{c2}) reduce to the standard commutation and
anticommutations relations
\begin{eqnarray}
\phi_i\phi_j^{\dagger}-\phi_j^{\dagger}\phi_i&=&\delta_{ij}\nonumber\\
\psi_i\psi_j^{\dagger}+\psi_j^{\dagger}\psi_i&=&\delta_{ij}, 
\end{eqnarray}
which, for $i,j=1,...N$, are covariant under $SU(N)$ transformations. 

\subsection{$SU_q(2)$-bosons}
For $N=2$, Equations (\ref{c1}) and (\ref{c2}) reduce to the set of equations
\begin{eqnarray}
\Phi_2\overline{\Phi}_2-q^2\overline{\Phi}_2\Phi_2&=&1 \label{1}\\
\Phi_1\overline{\Phi}_1-q^2\overline{\Phi}_1\Phi_1&=&1+(q^2-1)\overline{\Phi}_2\Phi_2
\\
\Phi_2\Phi_1&=&q\Phi_1\Phi_2\\
\Phi_2\overline{\Phi}_1&=&q\overline{\Phi}_1\Phi_2 \label{4}.
\end{eqnarray}

The operators $\Phi_j$ should not be confused with the so called $q$-boson oscillators. 
A set $(a_i,a_i^{\dagger})$ of $q$-bosons are defined, by the
relations \cite{Mf,B}
\begin{equation}
a_i a_i^{\dagger}-q^{-1}a_i^{\dagger}a_i=q^N ,\;\;\;
[a_i,a_j^{\dagger}]=0=[a_i,a_j],\label{q1}
\end{equation}
where $N|n\rangle=n|n\rangle$.  Several variations of Equation (\ref{q1})
are common in the literature.  By taking two sets of $q$-bosons, as defined in
Equation (\ref{q1}), it has been shown \cite{Mf,Ng}
that the operators 
\begin{equation}
J_+=a_2^{\dagger}a_1\;,\;\;J_-=a_1^{\dagger}a_2\;,\;\;
2J_3=N_2-N_1,
\end{equation}
provide a realization of the quantum Lie algebra $su_q(2)$
\begin{equation}
[J_3,J_{\pm}]=\pm J_{\pm}\;,\;\;\;\;
[J_-,J_-]=[2J_3].
\end{equation}

The main distinction between the $\Phi_j$ operators with $q$-bosons 
is that, in
contrast to Equations (\ref{1})-(\ref{4}),
the algebraic relations in Equation (\ref{q1}) are not covariant under the action
of the $SU_q(2)$ quantum group matrices.  Several studies on the thermodynamics of $q$-bosons, and similar
operators called quons \cite{G}, have been published \cite{TH,An}.   The work devoted to the thermodynamics of $q$-oscillators  represents a study
on the consequences of modifying  boson commutators, according to Equation (\ref{q1}) and
its different versions, and no new symmetry is imposed.  On the other hand, the algebraic relations in
Equations (\ref{1})-(\ref{4}), with the models discussed in this paper, 
address the implications that result of imposing
quantum group invariance in a thermodynamic system.

The simplest Hamiltonian written in terms of the operators $\Phi_j$ is given by
Hamiltonian with two species.  It is simply written as
\begin{equation}
{\cal H}_B=\sum_\kappa \varepsilon_\kappa({\cal N}_{1,\kappa}+{\cal N}_{2,\kappa}),\label{H}
\end{equation}
where $[\overline{\Phi}_{i,\kappa},\Phi_{\kappa',j}]=0$ for
$\kappa\neq\kappa'$ and ${\cal N}_{i,\kappa}=\overline{\Phi}_{i,\kappa}\Phi_{i,\kappa}$.
For a given $\kappa$ the $SU_q(2)$ bosons are written 
in terms of boson operators $\phi_{i}$
and $\phi_{i}^{\dagger}$ with usual commutation relations:
$[\phi_i,\phi_j^{\dagger}]=\delta_{ij}$ as follows
\begin{eqnarray}
\Phi_j&=&(\phi_j^\dagger)^{-1} \{N_j\}q^{N_{j+1}},\\
\nonumber\\
\overline{\Phi}_j&=&\phi_j^\dagger q^{N_{j+1}}, \;\;\; j=1,2
\end{eqnarray}
leading to the  interacting boson Hamiltonian
\begin{eqnarray}
{\cal H}_B&=&\sum_{\kappa}\varepsilon_{\kappa}\{N_{1,\kappa}+N_{2,\kappa}\},\nonumber\\
&=&\sum_{\kappa}\frac{\varepsilon_{\kappa}}{q^2-1}\sum_{m=1}^{\infty}
\frac{2^m \ln^mq}{m!}\left(N_{1,\kappa}+N_{2,\kappa}\right)^m,
\end{eqnarray}
where $N_{i,\kappa}$ is the ordinary boson number operator and the bracket
$\{x\}=(1-q^{2x})/(1-q^2)$. Equation (\ref{H}) becomes the standard free boson hamiltonian at $q=1$.

 The grand partition function ${\cal Z}_B$ is given by
\begin{equation}
{\cal Z}_B=\prod_\kappa\sum_{n=0}^{\infty}\sum_{m=0}^{\infty}
e^{-\beta\varepsilon_\kappa\{n+m\}}e^{\beta\mu(n+m)},\label{Zb}
\end{equation}
which after rearrangement simplifies to the Equation
\begin{equation}
{\cal Z}_B=\prod_\kappa\sum_{m=0}^{\infty} (m+1)e^{-\beta\varepsilon_{\kappa}\{m\}}z^m,\label{Z}
\end{equation}
where $z=e^{\beta\mu}$ is the fugacity. 
\subsection {$SU_q(2)$-fermions}
For $SU_q(2)$ fermions, Equations (\ref{c1}) and (\ref{c2}) become
\begin{eqnarray}
\{\Psi_{2},\overline{\Psi}_{2}\}&=&1 \label{f0}\\
\{\Psi_{1},\overline{\Psi}_{1}\}&=&1 - (1-q^{-2})\overline{\Psi}_{2}\Psi_{2}\label{f1}\\ 
\Psi_{1}\Psi_{2}&=&-q \Psi_{2}\Psi_{1}\\ 
\overline{\Psi}_{1}\Psi_{2}&=&-q \Psi_{2}\overline{\Psi}_{1}\\
\{\Psi_{1},\Psi_{1}\}&=&0=\{\Psi_{2},\Psi_{2}\}, \label{01}
\end{eqnarray}
The corresponding Hamiltonian in terms of
the operators $\Psi_i$ is simply 
\begin{equation}
{\cal H}_F=\sum_{\kappa}^{}\varepsilon_{\kappa}({\cal M}_{1,\kappa}+
{\cal M}_{2,\kappa}),\label{h}
\end{equation}
where ${\cal M}_{i\kappa}=\overline{\Psi}_{i,\kappa}\Psi_{i,\kappa}$
and $\{\overline{\Psi}_{\kappa,i},\Psi_{\kappa',j}\}=0$ for $\kappa\neq\kappa'$.
From Equation (\ref{01}) we see that the occupation numbers are restricted to $m=0,1$ and 
therefore $SU_q(N)$ fermions satisfy the Pauli exclusion principle. 
For a given $\kappa$ a normalized state is simply written
\begin{equation}
\overline{\Psi}_2^n\overline{\Psi}_1^m|0>,\;\;\; n,m=0,1 \label{s}
\end{equation}
 and the ${\cal M}_i$  operator satisfies
\begin{equation}
[{\cal M}_2,\Psi_1]=0 ,
\end{equation}
and
\begin{equation}
{\cal M}_1\Psi_2-q^2\Psi_2{\cal M}_1=0.
\end{equation}
The grand partition function is given by
\begin{equation}
{\cal Z}_F= Tr\; e^{-\sum_\kappa \varepsilon_\kappa({\cal M}_{1,\kappa}+{\cal M}_{2,\kappa})}
e^{\beta\mu(M_{1,\kappa}+M_{2,\kappa})},
\end{equation}
where $M_{i,\kappa}=\psi_{i,\kappa}^{\dagger}\psi_{i,\kappa}$ 
are the standard fermion number operators, and the trace is taken with
respect to the states in Equation (\ref{s}).  Since the pair $\Psi_2,\overline{\Psi}_2$
satisfies standard anticommutation relations we can identify it without 
any loss of generality with a fermion pair $\psi_2,\psi_2^\dagger$. In addition,
from Equations (\ref{f1}) and (\ref{01}) we see that the operator $\Psi_1(\overline{\Psi}_1)$
is a function of the operator $\psi_1(\psi_1^{\dagger})$ times a function of
 ${\cal M}_2$. A simple check shows that Equations (\ref{f0})-(\ref{01})
are consistent with the following representation of $\Psi_i$ operators
in terms of fermion operators $\psi_j$ 
\begin{eqnarray}
\Psi_2&=&\psi_2\\
\overline{\Psi}_2&=&\psi_2^{\dagger}\\
\Psi_1&=&\psi_1\left(1+(q^{-1}-1)M_2\right)\\
\overline{\Psi}_1&=&\psi_1^{\dagger}\left(1+(q^{-1}-1)M_2\right),
\end{eqnarray}
and according to Equations (\ref{c1}) and (\ref{c2}) this result easily
generalizes for arbitrary $N$ to
\begin{equation}
\Psi_m=\psi_m\prod_{l=m+1}^{N}\left(1+(q^{-1}-1)M_l\right),
\end{equation}
and similarly for the adjoint equation.

The grand partition function ${\cal Z}_F$ is written
\begin{eqnarray}
{\cal Z}_F&=&\prod_\kappa\sum_{n=0}^1\sum_{m=0}^1e^{-\beta\varepsilon_\kappa(n+m
-(1-q^{-2})mn}e^{\beta\mu(n+m)}\nonumber\\
&=&\prod_\kappa \left(1+2e^{-\beta(\varepsilon_\kappa-\mu)}+e^{-\beta\left
(\varepsilon_\kappa(q^{-2} +1)-2\mu\right)}\right) \label{ZF}
\end{eqnarray}
which for $q=1$ becomes the square of a single fermion type grand partition
function.  From Equation (\ref{ZF}) we see  that the original Hamiltonian becomes 
 the interacting 
Hamiltonian
\begin{equation}
H_F=\sum_\kappa \varepsilon_\kappa\left(M_{1,\kappa}+M_{2,\kappa}+(q^{-2}-1)
M_{1,\kappa}M_{2,\kappa}\right).\label{HF}
\end{equation}
Therefore the parameter $q\neq 1$ mixes the two
degrees of freedom in a nontrivial way through a quartic term in the Hamiltonian.

It is interesting to remark the distinction between $SU_q(2)$-fermions
with the so called $q$-fermions $b_i$ and $b_i^\dagger$. The $q$-fermionic
 algebra was introduced in \cite{Ng}
\begin{eqnarray}
bb^\dagger+qb^\dagger b&=&q^{N_q}\label{qf1}\\
b^\dagger b&=&[N_q]\\
bb^\dagger&=&[1-N_q]\\
b^2=&0&=b^{\dagger 2}\label{qf4},
\end{eqnarray}
where the bracket $[x]=\frac{q^x-q^{-x}}{q-q^-1}$ and the number operator
$N_q|n\rangle=n|n\rangle$ with $n=0,1$. The algebra in Equations (\ref{qf1})-(\ref{qf4}) is not invariant under
quantum group transformations.
\section{Scalar Curvature}\label{Scalar}
The partition functions in Equations (\ref{Z}) and (\ref{ZF}) have exponential form and therefore we can simply define the metric
according to Ref. \cite{JM} as
\begin{equation}
g_{\alpha\gamma}=\frac{\partial^2\ln Z}{\partial\beta^{\alpha}\partial\beta^{\gamma}},\label{g}
\end{equation}
in the two-dimensional parameter space with coordinates $\beta^1=\beta$ and $\beta^2=-\beta\mu$. Once the metric is
obtained, the scalar curvature follows from the basic definitions
\begin{equation}
R=\frac{2}{det g}R_{1212},
\end{equation}
where $det g=g_{11}g_{22}-g_{12}g_{12}$ and the non-vanishing part of the curvature tensor $R_{\alpha\beta\gamma\lambda}$ is given in terms of the
Christoffel symbols $\Gamma_{\alpha\beta\gamma}=\frac{1}{2}\frac{\partial g_{\alpha\beta}}{\partial \beta^{\gamma}}$ as follows
\begin{equation}
R_{\alpha\beta\gamma\lambda}=g^{\eta\theta}\left(\Gamma_{\eta\alpha\lambda}\Gamma_{\theta\beta\gamma}-\Gamma_{\eta\alpha\gamma}\Gamma_{\theta\beta\lambda}\right).
\end{equation}
The calculation of $R$ simply reduces to solve
$$ R=\frac{1}{2 (detg)^2}\left|\begin{array}{ccc} g_{11} & g_{22} & g_{12}  \\ g_{11,1} & g_{22,1} & g_{21,1} \\
g_{11,2} & g_{22,2}& g_{21,2} \end{array}\right|,$$
where $g_{\alpha\beta,\lambda}=\partial g_{\alpha\beta}/\partial\beta^{\lambda}$.
\subsection{Quantum Group Bose-Einstein case}
From Equations (\ref{Z}) and (\ref{g}) the components of the metric  can be written  in three and two spatial dimensions as follows
\begin{eqnarray}
g_{11}&=&C_1\frac{\beta^{-2}}{\lambda^D}\int_0^{\infty}x^{\nu}\ln f dx=C_1\frac{\beta^{-2}}{\lambda^D}a_{\nu}\nonumber,\\
g_{12}&=-&C_2\frac{\beta^{-1}}{\lambda^D}\int_0^{\infty} x^{\nu}\frac{f'}{f} dx=C_2\frac{\beta^{-1}}{\lambda^D}b_{\nu},\\
g_{22}&=&C_3\frac{1}{\lambda^D}\int_0^{\infty} x^{\nu}\left(\frac{f''}{f}-\frac{f'^2}{f}\right)dx=C_3\frac{1}{\lambda^D}c_{\nu},\nonumber
\end{eqnarray}
where we defined a new variable $x=\beta p^2/2m$, $\nu=\frac{D-2}{2}$, the coefficients are 
$C_1=\frac{15V}{2\sqrt{\pi}}$, $C_2=\frac{3V}{\sqrt{\pi}}$, $C_3=\frac{2V}{\sqrt{\pi}}$ for $D=3$, $C_1=2A$, $C_2=A=C_3$ for $D=2$, $\lambda$ is the thermal wavelength and $f'=\frac{\partial f}{\partial\gamma}$. The function 
\begin{equation}
f=1+\sum_{m=1}(m+1)e^{-x\{m\}}z^m.
\end{equation}
After taking the corresponding derivatives we find that the scalar curvature is given for $D=3$
\begin{equation}
R=\frac{5\sqrt{\pi}\lambda^3}{V}\left(\frac{c_{\nu}b^2_{\nu}+a_{\nu}d_{\nu}b_{\nu}-2a_{\nu}c^2_{\nu}}{(5a_{\nu}c_{\nu}-3b^2_{\nu})^2}\right),\label{R3}
\end{equation}
and for $D=2$ 
\begin{equation}
R=2\frac{\lambda^2}{A}\left(\frac{c_{\nu}b^2_{\nu}+a_{\nu}d_{\nu}b_{\nu}-2a_{\nu}c^2_{\nu}}{(2a_{\nu}c_{\nu}-b^2_{\nu})^2}\right),\label{R2}
\end{equation}
where $d_{\nu}=\partial c_{\nu}/\partial\gamma$ and the  values of $\nu=1/2$ and $\nu=0$ for $D=3$ and $D=2$ respectively. In order to compare our results with the standard $q=1$ case and the fact that $SU_q(2)$ operators are doublets we
have multiplied our scalar curvature results by a factor of $2$. In general, for $SU_q(N)$ the scalar curvature will have an additional factor of $1/N$ that will have to be removed so we can  compare the results  with the scalar curvature for $q=1$ and $N=1$.
Figures 1 and 2 display the results of a numerical calculation of Equation (\ref{R3}) for the scalar curvature of a QGB system in $D=3$.
Figure 1 shows that for $0<q\leq 1$ the curvature is positive for all $z$ values, and therefore the system is bosonic. In addition, the system becomes  more stable than the standard Bose-Einstein case at higher values of $z$. For $1<q\approx 1.2$ the system exhibits anyonic behavior as a function of $z$ and becomes  purely fermionic for all values of $z$ and $q>1.2$. 
Figure 2 is a graph of the scalar curvature for $D=3$ as a function of the quantum group parameter $q$ at $z=0.2,0.5, 0.8$.
The graph shows a maximum instability that moves nearer the value $q=1$ as the fugacity increases its value. The system mimics a classical behavior in the interval $1<q<1.27$. In particular, for $q>>1$
 the scalar curvature remains negative and  approximately constant.
Figures 3 and 4 show graphs of the scalar curvature for $D=2$. Figure 3 displays the scalar curvature for $q=0.5,1,1.15$ as a function of $z$. For $0<q<1$ the system is bosonic and more stable at $z\approx 1$ than the standard case $q=1$. For $q>1$ the system goes from bosonic to  fermionic and becomes bosonic becoming more unstable as $z$ approaches the value of $1$. Figure 4 shows the change in the scalar curvature for three values of $z$ as a function of the quantum parameter $q$. For the values 
$0<z\approx 0.73$ and $q>1$ the system becomes fermionic increasing its instability with the value of $z$ with $R$ approximately constant for $q>>1$. For $z>0.73$, the system goes from $bosonic\rightarrow fermionic\rightarrow bosonic$ behavior keeping $R$ a constant positive value   for $q>>1$. In particular, the system is more unstable at higher values of $z$.  For $0.2<z<0.8$ the scalar curvature vanishes in the interval $1.12<q<1.3$.
\begin{figure}
\begin{center}
\epsfig{file= 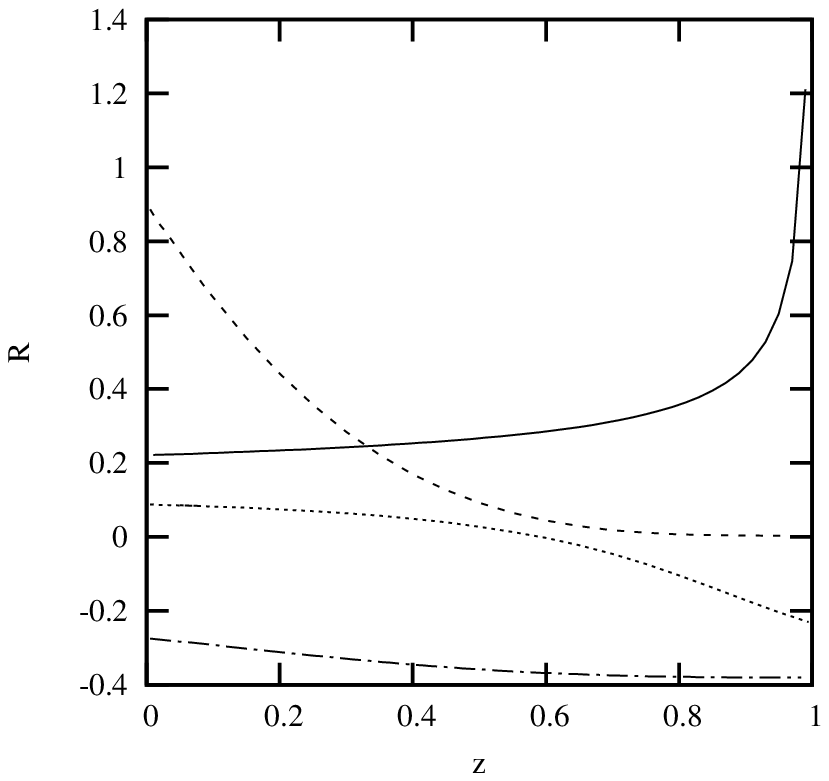,bbllx=50pt,bblly=120pt,bburx=430pt,bbury=250pt}
\end{center}
\caption[]{The scalar curvature $R$, in units of $\lambda^3V^{-1}$, as a function of the fugacity $z$ for quantum group bosons at $D=3$ and constant $\beta$ for the cases of $q=1$ (solid
line), $q=0.5$ (dashed line) and $q=1.15$ (dotted line) and $q=2$ (dashed-dotted line).}
\end{figure}
\begin{figure}
\begin{center}
\epsfig{file= 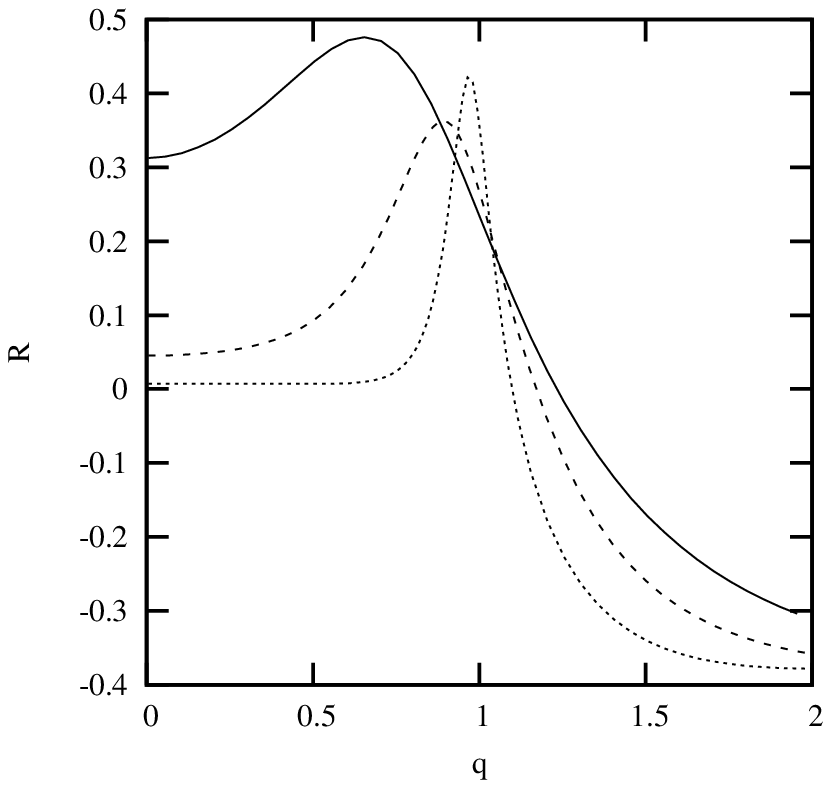,bbllx=50pt,bblly=120pt,bburx=430pt,bbury=250pt}
\end{center}
\caption[]{The scalar curvature $R$  for quantum group bosons at $D=3$, in units of $\lambda^3V^{-1}$, as a function of the parameter $q$ and values for
the fugacity $z=0.2$ (solid line),   $z=0.5$ (dashed line) and    $z=0.8$ (dotted line).}
\end{figure}
\begin{figure}
\begin{center}
\epsfig{file= 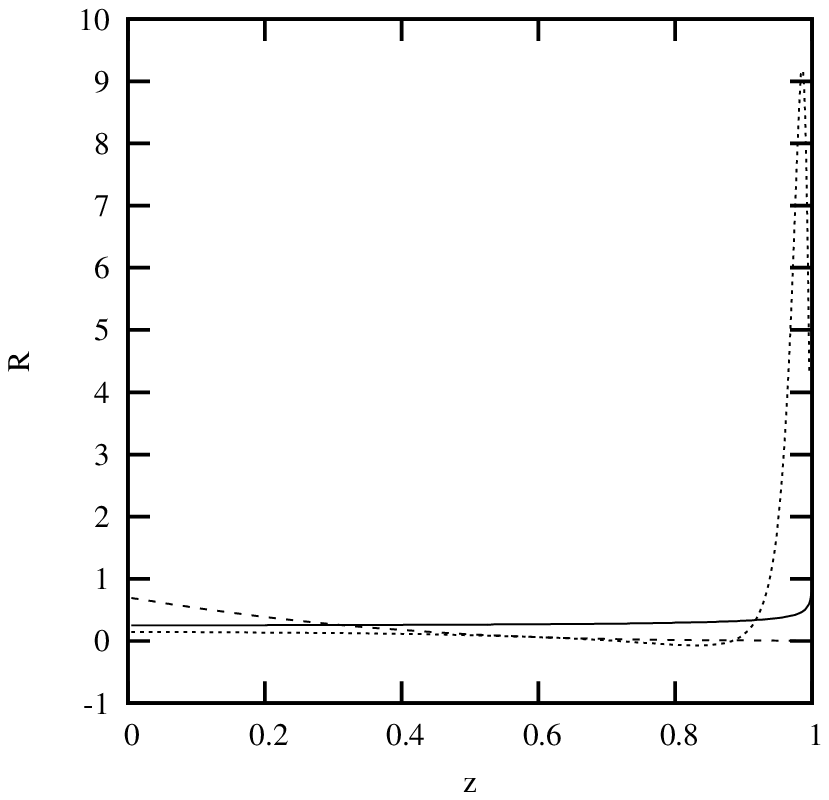,bbllx=50pt,bblly=120pt,bburx=430pt,bbury=250pt}
\end{center}
\caption[]{The scalar curvature $R$, in units of $\lambda^2A^{-1}$, as a function of the fugacity $z$ for quantum group bosons bosons at $D=2$ and  constant $\beta$ for the cases of $q=1$ (solid
line), $q=0.5$ (dashed line) and $q=1.15$ (dotted line).}
\end{figure}
\begin{figure}
\begin{center}
\epsfig{file= 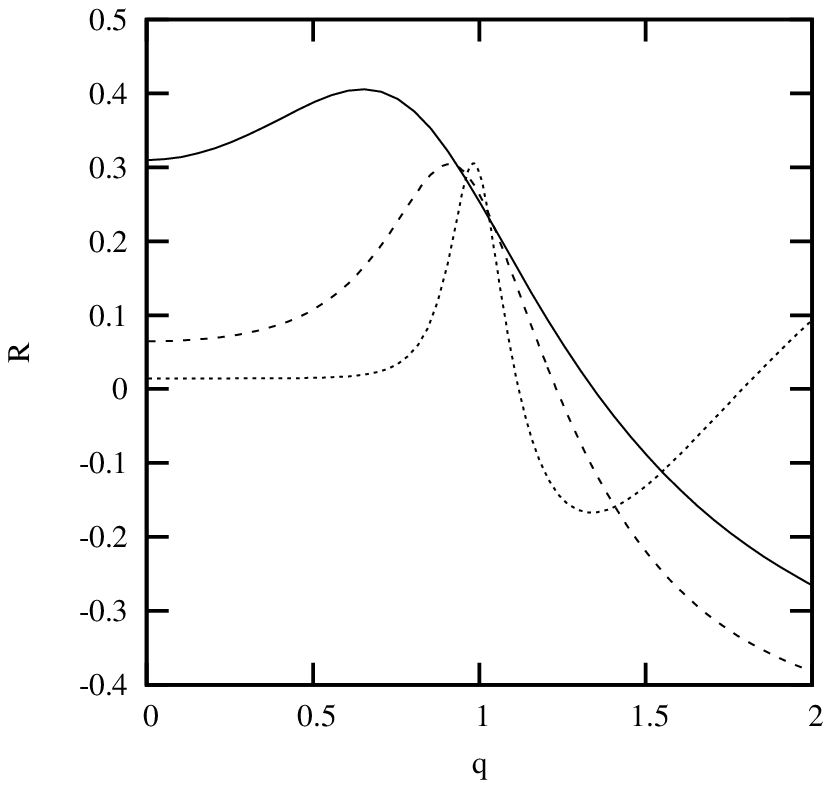,bbllx=50pt,bblly=120pt,bburx=430pt,bbury=250pt}
\end{center}
\caption[]{The scalar curvature $R$  for quantum group bosons at $D=2$, in units of $\lambda^2A^{-1}$, as a function of the parameter $q$ and values for
the fugacity $z=0.2$ (solid line),   $z=0.5$ (dashed line) and   $z=0.8$ (dotted line).}
\end{figure}
\begin{figure}
\begin{center}
\epsfig{file= 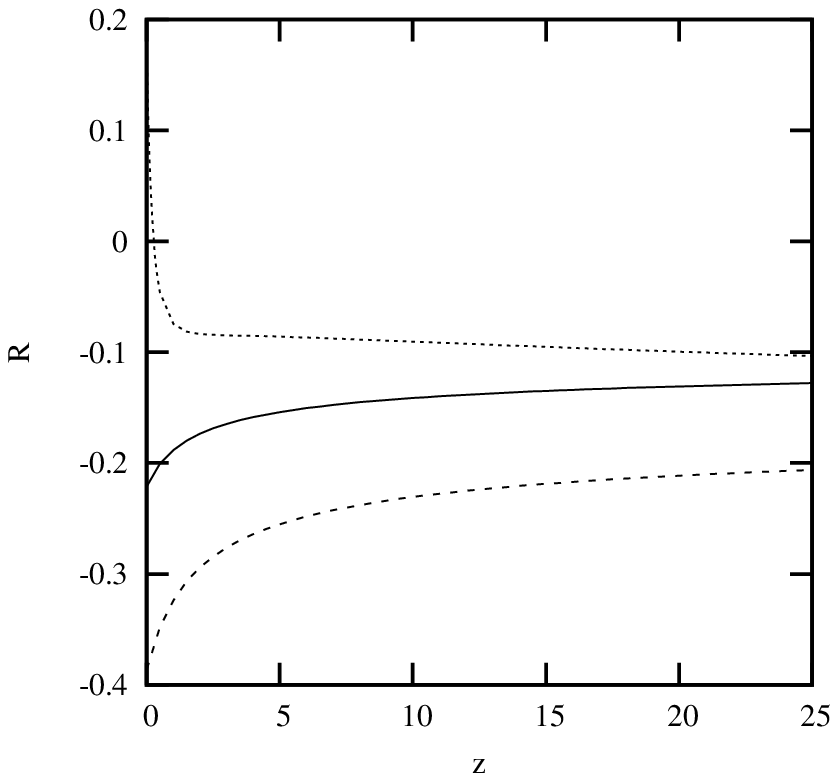,bbllx=50pt,bblly=120pt,bburx=430pt,bbury=250pt}
\end{center}
\caption[]{The scalar curvature $R$  for quantum group fermions at $D=3$, in units of $\lambda^3V^{-1}$, as a function of $z$ at constant $\beta$  for the values
 $q=1$ (solid line),   $q=0.5$ (dashed line),  and $q=10$ (dotted line).}
\end{figure}
\begin{figure}
\begin{center}
\epsfig{file= 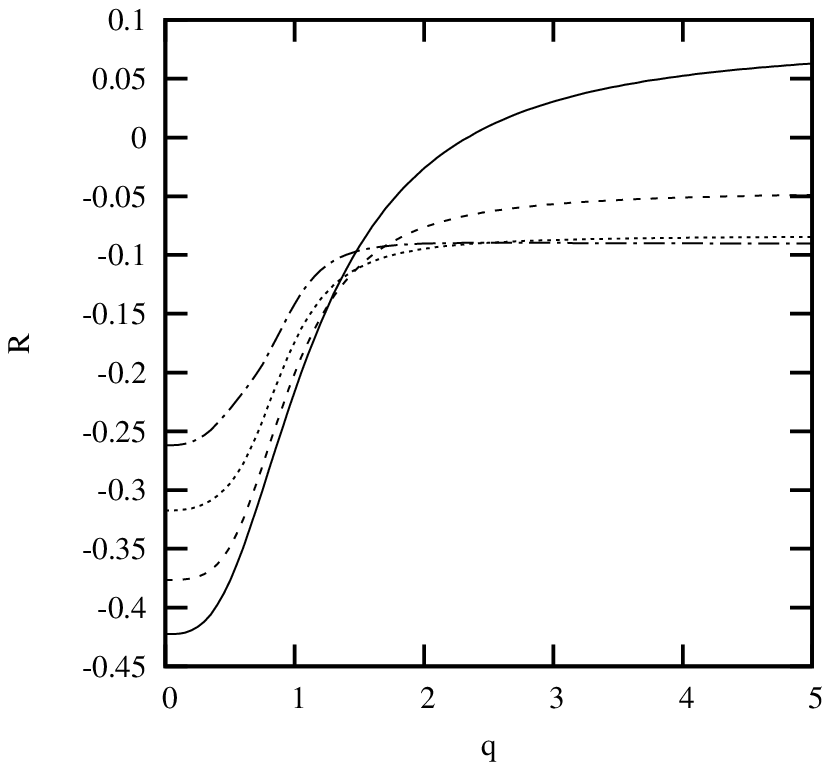,bbllx=50pt,bblly=120pt,bburx=430pt,bbury=250pt}
\end{center}
\caption[]{The scalar curvature $R$  for quantum group fermions at $D=3$, in units of $\lambda^3V^{-1}$, as a function of the parameter $q$ and values for
the fugacity $z=0.1$ (solid line),   $z=0.5$ (dashed line),   $z=2$ (dotted line) and $z=10$ (dashed-dotted line).}
\end{figure}
\begin{figure}
\begin{center}
\epsfig{file= 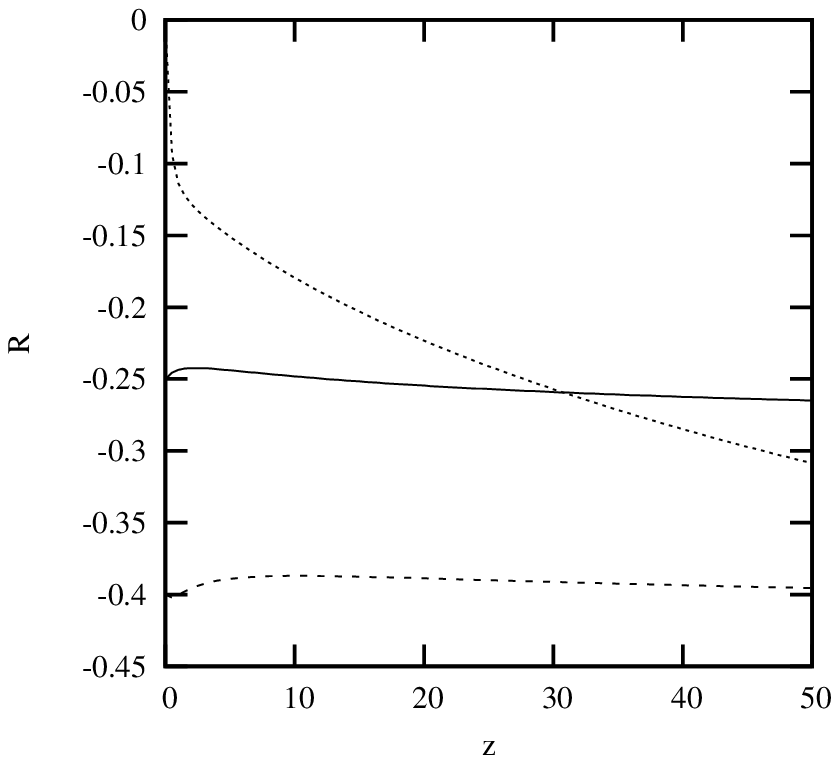,bbllx=50pt,bblly=120pt,bburx=430pt,bbury=250pt}
\end{center}
\caption[]{The scalar curvature $R$  for quantum group fermions at $D=2$, in units of $\lambda^2A^{-1}$, as a function of $z$ at constant $\beta$  for the values
 $q=1$ (solid line),   $q=0.5$ (dashed line)  and $q=10$ (dotted line).}
\end{figure}
\begin{figure}
\begin{center}
\epsfig{file= 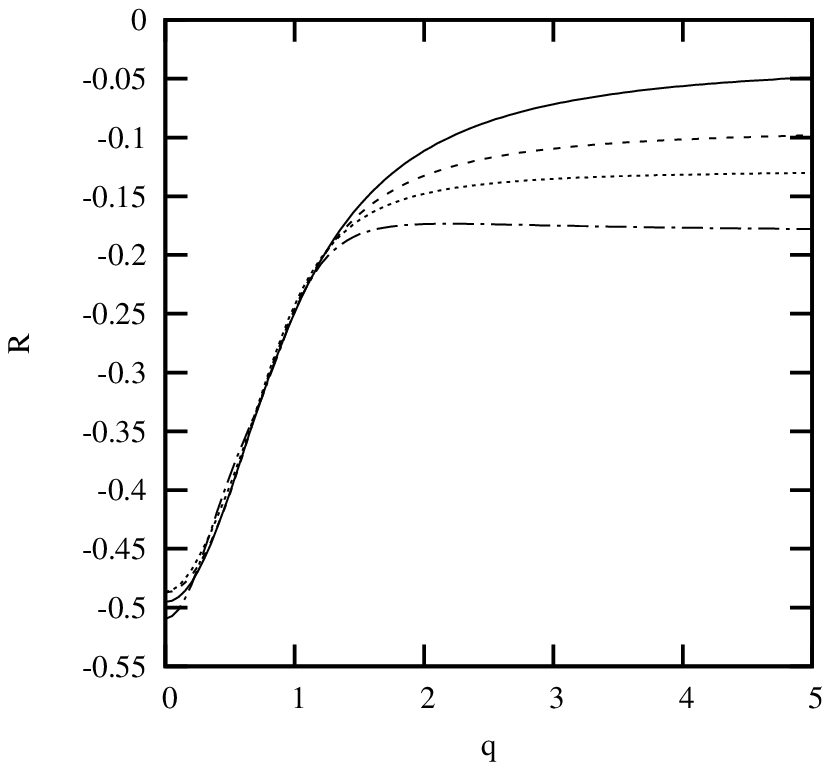,bbllx=50pt,bblly=120pt,bburx=430pt,bbury=250pt}
\end{center}
\caption[]{The scalar curvature $R$  for quantum group fermions at $D=2$,in units of $\lambda^2A^{-1}$, as a function of the parameter $q$ and values for
the fugacity $z=0.1$ (solid line),   $z=0.5$ (dashed line),   $z=2$ (dotted line) and $z=10$ (dashed-dotted line).}
\end{figure}
\subsection{Quantum Group Fermi-Dirac case}
From Equation (\ref{ZF}) we obtain for the components of the metric
\begin{eqnarray}
g_{11}&=&C_1\frac{\beta^{-2}}{\lambda^D}\int_0^{\infty}x^{\nu}\ln h\;  dx=C_1\frac{\beta^{-2}}{\lambda^D}{\tilde a}_{\nu}\nonumber,\\
g_{12}&=-&C_2\frac{\beta^{-1}}{\lambda^D}\int_0^{\infty} x^{\nu}\frac{h'}{h} dx=C_2\frac{\beta^{-1}}{\lambda^D}{\tilde b}_{\nu},\\
g_{22}&=&C_3\frac{1}{\lambda^D}\int_0^{\infty} x^{\nu}\left(\frac{h''}{h}-\frac{h'^2}{h}\right)dx=C_3\frac{1}{\lambda^D}{\tilde c}_{\nu},\nonumber
\end{eqnarray}
where the function $h$ is given by
\begin{equation}
h=1+2e^{-x}z+e^{-(q^{-2}+1)x}z^2,
\end {equation}
leading to equations identical to Equations (\ref{R3}) and (\ref{R2}) with the replacements  of $a_{\nu}$, $b_{\nu}$ and $c_{\nu}$ by
${\tilde a}_{\nu}$, ${\tilde b}_{\nu}$ and ${\tilde c}_{\nu}$ respectively.
Figures 5 and 6 show the scalar curvature for a QGF system at $D=3$. Figure 5 shows that for $q<1$ the system is fermionic
and more unstable than the $q=1$ case. For $q=10$ the system is bosonic for small values of $z$ becoming fermionic and more stable at higher values of $z$ as compared with the $q=1$ case. For $z>25$ the scalar curvature is approximately constant. Figure 6 is a graph of $R$ vs. $q$ for the values $z=0.1,0.5,2,10$. For $z=0.1$ the system becomes bosonic for $q\geq 2.3$. This QGF system is more stable 
than the $q=1$ system when $q<1$ and it is more unstable than the $q=1$ system when $q>1$.
Figures 7 and 8 display the results for a QGF system for $D=2$. Figure 7 shows that for $q<1$ a QGF system is less stable than for $q=1$, and for $q>1$ 
is more stable  at lower values of $z$. Figure 8 shows the results of $R$ vs. $q$ for four values of $z$.
For $q<1$ the scalar curvature is almost independent of $q$. However, for $q>1$ there is more stability at lower values of $z$. Quantum group fermions do not exhibit anyonic behavior at $D=2$.

For low $z$ we can simply find the value of $q$ at which the system changes from bosonic to fermionic and viceversa. 
For the case of QGF systems for $D=3$ at high temperatures (very low $z$) we obtain \cite{MRU2} from Equation (\ref{ZF})
at second order in $z$
\begin{equation}
\ln{Z_F}=\frac{2^{7/2}\pi^{3/2}V}{\lambda^3}\left(\frac{z}{2}-\alpha(q) z^2+...\right),
\end{equation} 
where 
\begin{equation}
\alpha(q)=\frac{1}{2}\left(\frac{1}{2^{3/2}}-\frac{1}{2(q^{-2}+1)^{3/2}}\right).
\end{equation}
From the total average number of particles
\begin{eqnarray}
<N>&=&\frac{1}{\beta}\frac{\partial \ln{Z_F}}{\partial \mu}\nonumber\\
&=&\frac{2^{7/2}\pi^{3/2}V}{\lambda^3}\left(\frac{z}{2}-\alpha(q)\frac{8<N>^2\beta^3}{\lambda^2\pi}\right)
\end{eqnarray}
we obtain at second order in $<N>$
\begin{equation}
z=\frac{\lambda^3}{\pi^{3/2}2^{5/2}}\frac{<N>}{V}+\frac{\alpha(q)\lambda^6}{\pi^22^4}\left(\frac{<N>}{V}\right)^2.
\end{equation}
For $q<1.96$ the function $\alpha(q)$ is positive and therefore the QGF  behaves as a fermion system, and for $q>1.96$
the function $\alpha(q)$ becomes negative and therefore QGF become bosonic.
For QGB systems we have
\begin{eqnarray}
z&=& \frac{1}{2}\lambda^3\frac{<N>}{V}+\delta(q)\lambda^6\left(\frac{<N>}{V}\right)^2,\\
z&=&\frac{1}{2}\lambda^2\frac{<N>}{A}+\eta(q)\lambda^4\left(\frac{<N>}{A}\right)^2
\end{eqnarray}
with the corresponding parameters  
\begin{eqnarray}
\delta&=-&\frac{1}{4}\left(\frac{3}{(1+q^2)^{3/2}}-\frac{1}{\sqrt{2}}\right),\nonumber\\
\eta&=-&\frac{2-q^2}{4(1+q^2)},
\end{eqnarray}
for $D=3$ and $D=2$ respectively. For $D=3$ and $q>1.27$ the function $\delta(q)>0$ and for $D=2$ and $q>\sqrt{2}$ the function $\eta(q)>0$  and the QGB systems become fermionic. 

Certainly, the advantage of the calculation of the scalar curvature is that it gives
a description of the anyonic behavior in the whole range of $z$ values, which means high and low temperatures.

We summarize the sign of the scalar curvature for $D=3$ and $D=2$ for $z\approx 0$ in the following two tables:

$$D=3$$
\begin{center}
\begin{tabular}{|l|l|l|r||}\hline
QGB & $q<1.27$ & $R>0$ \\ \cline{2-3}
& $q>1.27$ & $R<0$\\ \hline
QGF & $q<1.96$ & $R<0$ \\ \cline{2-3} 
& $q>1.96$ & $R>0$ \\ \hline
\end{tabular}
\end{center}

$$D=2$$
\begin{center}
\begin{tabular}{|l|l|l|r||}\hline
QGB & $q<\sqrt{2}$ & $R>0$ \\ \cline{2-3}
& $q>\sqrt{2}$ & $R<0$\\ \hline
QGF & $\forall q$ & $R<0$ \\ \hline 
\end{tabular}
\end{center}

\begin{figure}
\epsfig{file= 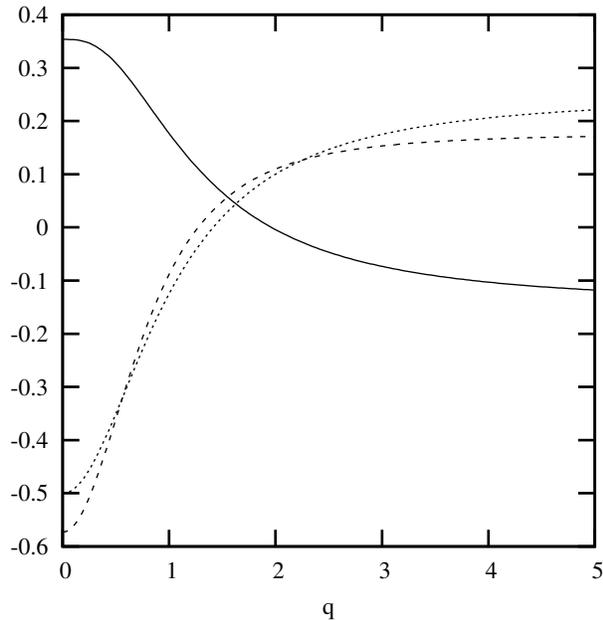,bbllx=50pt,bblly=120pt,bburx=430pt,bbury=250pt}
\caption[]{The virial coefficients $\alpha(q)$ (solid line), $\delta(q)$ (dashed line) and $\eta(q)$ (dotted line) for
QGF in $D=3$, QGB in $D=3$ and QGB in $D=2$ respectively}
\end{figure}
\vspace{0.3in}

Figure 9 displays the dependence of the virial coefficients $\alpha(q)$, $\delta(q)$ and $\eta(q)$ on the quantum group parameter.
For the case of QGF in $D=2$ the sign of the virial coefficient does not change $\forall q$ and no anyonic behavior occurs. 
\section{Conclusions} \label{Conc}
In this manuscript we have calculated the scalar curvature $R$ for systems with quantum group symmetry. The values of the scalar curvature tell us about the corelations and therefore the stability of the system. In addition, since $R$ is positive (negative) for bosons (fermions) these calculations give us a more detailed picture about their anyonic behavior
than a calculation of the virial coefficients could provide. There is no anyonic behavior for those values $0<q\leq 1$, and anyonic behavior occurs for QGB systems in $D=3$ and $D=2$ and QFG systems in $D=3$. In particular, for QGB systems
in $D=2$   the behavior goes from bosonic to fermionic and then bosonic near the value 
$z=1$ in the neighborhood of  $q=1.15$. In general, QGB are more unstable for $0<q<1$ and $z=0$ and more stable
near $z=1$ than  for $q=1$. For QGF systems in $D=3$ our results show more instability for $0<q<1$ and more
stability for $q>1$, while for $D=2$ the scalar curvature is almost $q$ independent for $0<q<1$ becoming more stable for larger values of $q$. The QGF system in $D=2$ is the only one that does not exhibit anyonic behavior at any temperature.
Although at this moment is still an open question the relevance that  quantum groups may have outside the theory of the inverse scattering method and  integrable models in the context of Yang-Baxter algebras, and  whether the interactions induced by the requirement of quantum group invariance are realizable in quantum statistical mechanics,
 this work  provides a theoretical framework to understand boson-fermion transmutation in two and three dimensions which could have implications in other areas of physics.

\end{document}